\documentclass[galaxies,review,accept,moreauthors,pdftex]{Definitions/mdpi}

\firstpage{1}
\pubvolume{9}
\issuenum{2}
\articlenumber{42}
\pubyear{2021}
\copyrightyear{2021}
\externaleditor{Academic Editors: Margo Aller, Jose L. Gómez and Eric Perlman} 
\datereceived{13 April 2021}
\dateaccepted{10 June 2021}
\datepublished{11 June 2021}
\hreflink{https://doi.org/10.3390/\linebreak galaxies9020042} 

\Title{Polarimetric Properties of Blazars  {Caught} by the WEBT}

\TitleCitation{Polarimetric Properties of Blazars {Caught} by the WEBT}

\Author{Claudia M. Raiteri
*\href{https://orcid.org/0000-0003-1784-2784}{\orcidicon} and Massimo Villata \href{https://orcid.org/0000-0003-1743-6946}{\orcidicon}}

\AuthorNames{Claudia M. Raiteri and Massimo Villata}

\AuthorCitation{Raiteri, C.M.; Villata, M.}

\address[1]{%
INAF, Osservatorio Astrofisico di Torino, via Osservatorio 20, I-10025 Pino Torinese, Italy; massimo.villata@inaf.it}

\corres{\hangafter=1 \hangindent=1.05em \hspace{-0.82em}Correspondence: claudia.raiteri@inaf.it}

\abstract{Active galactic nuclei come in many varieties.
A minority of them are radio-loud, and exhibit  {two opposite} prominent plasma jets extending from the proximity of the supermassive black hole up to megaparsec distances.
When one of the relativistic jets is oriented closely to the line of sight, its emission is Doppler beamed   and these objects show extreme variability properties at all wavelengths. These are called ``blazars''.
The unpredictable blazar variability, occurring on a continuous range of time-scales, from minutes to years, is most effectively investigated in a multi-wavelength context.
Ground-based and space observations together contribute to give us a comprehensive picture of the blazar emission properties from the radio to the $\gamma$-ray band.
Moreover, in recent years, a lot of effort has been devoted to the observation and analysis of the blazar polarimetric radio and optical behaviour, showing strong variability of both the polarisation degree and angle.
The Whole Earth Blazar Telescope (WEBT) Collaboration, involving many tens of astronomers all around the globe, has been monitoring several blazars since 1997. The results of the corresponding data analysis have contributed to the understanding of the blazar phenomenon, particularly stressing the viability of a geometrical interpretation of the blazar variability. We review here the most significant polarimetric results achieved in the WEBT studies.}

\begin{document}

\section{Introduction}

According to the commonly accepted scenario for active galactic nuclei (AGN), they are powered by a supermassive black hole (SMBH), which is fed by an accretion disc. Here, matter is thought to spiral around the SMBH, until it falls onto it, releasing a huge amount of energy. A fraction of AGN, around 10--20\%, are radio-loud, which classically means that they have a large ratio of the radio flux  {at 5 GHz} over the optical one  {in the $B$ band} $F_{\rm 5~GHz}/ F_B \ge 10$, e.g., \citep{urry1995}. These objects show radio jets coming out from the nuclear region,  {in a direction presumed to be perpendicular} to the accretion disc  {or along the SMBH rotation axis, which might not be aligned with the normal to the accretion disc}. The plasma moves at relativistic speed, at least in the inner jet regions,  {at parsec scales,} emitting synchrotron and inverse-Compton radiations.
If one of the two jets points towards the Earth, we see its emission Doppler beamed. As a consequence, the jet radiation appears blue-shifted (in contrast to the red-shift due to the expansion of the Universe), and enhanced by some power of the Doppler factor. Moreover, the variability time-scales  {shorten inversely to the Doppler factor.} The objects where this situation occurs are called ``blazars'' and show unpredictable strong variability at all wavelengths, from the radio to the $\gamma$-ray band, on a continuous range of time-scales. Fast variations, even on intraday scales, usually overlap with progressively wider oscillations on the time-scales of several weeks, months or years. In some cases, indications of quasi-periodicity have been found, as in the cases of the BL Lac objects OJ~287 \cite{sillanpaa1996} and PG~1553 + 113 \cite{ackermann2015}.

Actually, the blazar class includes a variety of objects that share many properties, but also present differences. A subclass of blazars is represented by the flat-spectrum radio quasars (FSRQs), which show strong emission lines in their spectrum,  {including broad lines produced by fast-moving gas clouds in the nuclear region, and narrow lines due to gas clouds moving more slowly in a more external region}  \citep{urry1995}.   {Moreover, FSRQs also show} a clear signature of the accretion disc emission in their spectral energy distribution (SED), a feature called ``big blue bump''. In contrast, the subclass of BL Lac objects includes sources whose spectra are characterised by the absence or extreme weakness of emission lines. BL Lac objects are then further subdivided into three classes of low-, intermediate-, and high-energy peaked BL Lacs, according to the position of their synchrotron peak in the~SED.

Blazar flux variations are usually accompanied by spectral variations. In general,  {in the optical band} BL Lac objects show a bluer-when-brighter trend, while in FSRQs, the presence of the big blue bump due to the accretion disc produces a redder-when-brighter trend, which may turn into a bluer-when-brighter behaviour in the brightest states \cite{villata2006}. Moreover, long-term trends have in some cases showed a less chromatic behaviour than the fast variations \cite{raiteri2021}.

In recent years, polarisation has acquired more and more importance in blazar studies, as it can add a fundamental piece of information in our understanding of the physical structure of blazar jets, and of their emission and variability mechanisms. Most blazar polarimetric observations are done in the radio and optical bands, with additional data acquired in the near-infrared band. In the near future, the {\it IXPE} space mission 
(\url{https://ixpe.msfc.nasa.gov/}, accessed on 10 June 2021) will allow us to study blazar polarimetry also at X-ray energies.

The Whole Earth Blazar Telescope (\url{https://www.oato.inaf.it/blazars/webt/}, accessed on 10 June 2021) (WEBT) \citep{villata2002,villata2004,villata2008} was created in 1997 with the aim of organising multiband monitoring campaigns on blazars to understand their variability properties. Several tens of astronomers from all over the world participated in the WEBT activities and to date still perform observations, mainly in the optical but also in the radio and near-infrared bands. The data analysis often involves supporting high-energy observations by satellites at UV+X-ray frequencies like {\it XMM-Newton} and {\it Swift}, at $\gamma$-ray energies like {\it Fermi}, or by ground-based Cherenkov telescopes like MAGIC, observing in the TeV energy domain.

\section{Polarimetric Signatures in Blazars}

The blazar synchrotron emission detected from radio to optical--{UV} bands is  {linearly} polarised, with variable polarisation degree $P$ and electric vector  {position} angle \linebreak~EVPA \cite{smith1996}. The behaviour of $P$, which is observed to reach values as high as $\geq$50\% in the optical, is sometimes found to correlate with the flux, increasing with the source brightness, but often it does not. The radio and optical EVPAs can show  {large} rotations \citep{ledden1979, aller1981,kikuchi1988,marscher2008,marscher2010,sasada2012,morozova2014,blinov2015,blinov2018} not necessarily linked to particular flux states or events. Indeed,  {large} rotations can also be obtained from stochastic processes like turbulence, including many plasma cells with the random orientation of the magnetic field  \cite{darcangelo2007,marscher2014,blinov2015,kiehlmann2016}. Moreover, it was shown by \cite{lyutikov2017} that when changes of the jet orientation occur, variations of $P$ and EVPA may look stochastic even if they are produced by deterministic processes. This makes the understanding of the physical scenario more challenging.

Various models have been proposed to explain the observed EVPA swings, including:
(i) a relativistic source accelerated by shock waves in a uniform magnetic field \citep{blandford1979};
(ii) a source in relativistic motion in a magnetic field with three-dimensional distribution \citep{bjornsson1982};
(iii) shock waves propagating in jets with non-axisymmetric magnetic field \citep{konigl1985} or helical magnetic fields \citep{larionov2013,zhang2014,zhang2016};
(iv) a jet with helical magnetic field and variable viewing \mbox{angle \citep{lyutikov2017};} and
(v) magnetic reconnection \citep{zhang2020}.

However, there are a couple of caveats that one should keep in mind before trying to interpret the observational results.
One caveat concerns the reliability of EVPA rotations~\cite{larionov2016b}. As known, there is a $\pm n \times 180^\circ$ ambiguity in the EVPA direction, and the reconstruction of the EVPA temporal behaviour is usually done by minimising the jump between subsequent data points. When this jump is around $90^\circ$, then the reconstruction becomes quite subjective and one can introduce false rotations.
{This problem can in principle be eliminated with a suitable observational sampling. In most cases, EVPA changes $\ge$$90^\circ$ take several days. However, fast rotations have been observed in the optical band, for which sub-day sampling is needed} \citep{kiehlmann2021}.
{Moreover, the presence of thermal plasma in the jet or in the immediate vicinity of the outflow, identified from the Faraday rotation effect observed in parsec-scale AGN jets, makes the situation even more complex.}

{In the optical band,} another caveat is the presence of  {unpolarised} thermal emission due to the accretion disc in the FSRQs where the big blue bump is clearly recognisable in the SED. For those cases, if we want to understand what happens in the jet, the {contribution of the unpolarised disc} must be subtracted from the total flux, and the polarisation degree must be corrected for its ``dilution'' effect as
\begin{equation}
P_{\rm jet}=(P_{\rm obs} \times F_{\rm obs})/(F_{\rm obs}-F_{\rm disc}),
\label{dilution}
\end{equation}
which requires an estimate of the disc contribution to the total flux. This dilution effect is what sometimes makes the observed polarisation much smaller than the jet one. In any case, the jet polarisation never reaches the theoretical maximum value predicted for a power-law electron distribution in a uniform magnetic field of
$P_0 = (\alpha+1)/(\alpha+5/3) \sim$ 0.69--0.75
(for an electron energy distribution
$N(E) \propto E^{-(2 \alpha + 1)}$
with typically $\alpha$ = 0.5--1). Indeed, depending on the symmetry of the magnetic field, the contributions from differently oriented field lines can partly eliminate each other.
We notice that  {an overestimation of} $F_{\rm disc}$ in Equation \eqref{dilution} may even lead to a violation of the above theoretical limit on $P_{\rm jet}$. Therefore, this limit puts a constraint on the assumed contribution for the thermal component.
A similar caveat holds for some close BL Lac objects, for which the dilution effect is played by the host galaxy. In this case, $P_{\rm jet}=(P_{\rm obs} \times F_{\rm obs})/(F_{\rm obs}-F_{\rm host})$.

\section{Polarimetric Results by the WEBT}

The WEBT studies have included polarimetry since 2008.
In  {the} analysis of the multifrequency behaviour of the well-known FSRQ 3C~279 during the period of 2006--2007, Ref. \cite{larionov2008} detected an impressive  {large} rotation (about $300^\circ$ in two months) of the polarisation angle of both the VLBA radio core at 43 GHz and the optical emission in the $R$ band.
This occurred during a remarkable fading phase of $\sim$3.5 mag in $\sim$100 d. Because of the smooth trend of the rotation,  a stochastic effect was ruled out, and the event was interpreted as the signature of a compressive feature moving along a helical magnetic field.

The multifrequency variability of another FSRQ, 4C~38.41, was studied by \cite{raiteri2012}.
They gave a geometric explanation to the lack of optical-radio correlation and to the changeable optical-$\gamma$ correlation. The interpretation involved an inhomogeneous bent jet, where emission at different frequencies come from distinct regions in the jet, which can have a different orientation with respect to the line of sight. In this scenario, the long-term flux variability at a given frequency is produced by changes in the Doppler factor \linebreak~$\delta=1/[\Gamma (1-\beta \cos \theta)]$ due to variations in the viewing angle $\theta$ of the corresponding jet emitting region. $\Gamma$ is the bulk Lorentz factor and $\beta$ is the plasma velocity in units of the speed of light.
The observed correlation between $P_{\rm jet}$ and $F_{\rm jet}$ was then explained by a shock-in-jet model,  {where} shock waves with variable  {degree of compression} $\eta$  {(representing the ratio between the lengths before and after the passage of the compressing features), travel} along the bent jet, with consequent:
\begin{equation}
P \approx P_0 \, { {(1-\eta^{-2}) \sin^2\theta'}\over {2-(1-\eta^{-2}) \sin^2\theta'}},
\label{hughes}
\end{equation}
where, because of relativistic aberration the viewing angle in the rest frame, $\theta'$ transforms into the  {angle $\theta$ in the observer's frame} according to the equation
$\sin \theta'=\delta \, \sin \theta$.
\mbox{{Equation} \eqref{hughes}}  {is a reformulation of the expression derived by} \citet{hughes1985}  {for the degree of polarisation resulting after an initially random magnetic field is compressed by \mbox{a shock.}}

The same geometrical scenario was applied by \cite{raiteri2013} to the flux and polarisation behaviour of BL Lacertae, the eponym of the BL Lac class.
{During the period 2011--2012, this source} was observed in a remarkable optical and $\gamma$-ray flaring state, after a long period of moderate activity.
During the low state, the average polarisation degree was larger than in the following bright phase, and the EVPA was roughly stable around $\sim$$15^\circ$, while during the flaring phase it underwent large oscillations.
The prediction of a simple jet model with helical magnetic field and variable viewing angle was compared to the shock-in-jet model previously applied to 4C~38.41, and both gave the same results for a reasonable choice of model parameters. However, the shock-in-jet interpretation showed to be superior, as the possibility to involve shock waves of variable strength allowed to derive the trend of $\eta$ that can fully reproduce the  {long-term behaviour of $P$ in time, represented by a cubic spline interpolation through the 60-day binned observed polarisation curve}.

\begin{figure}[H]
\includegraphics[width=10.5 cm]{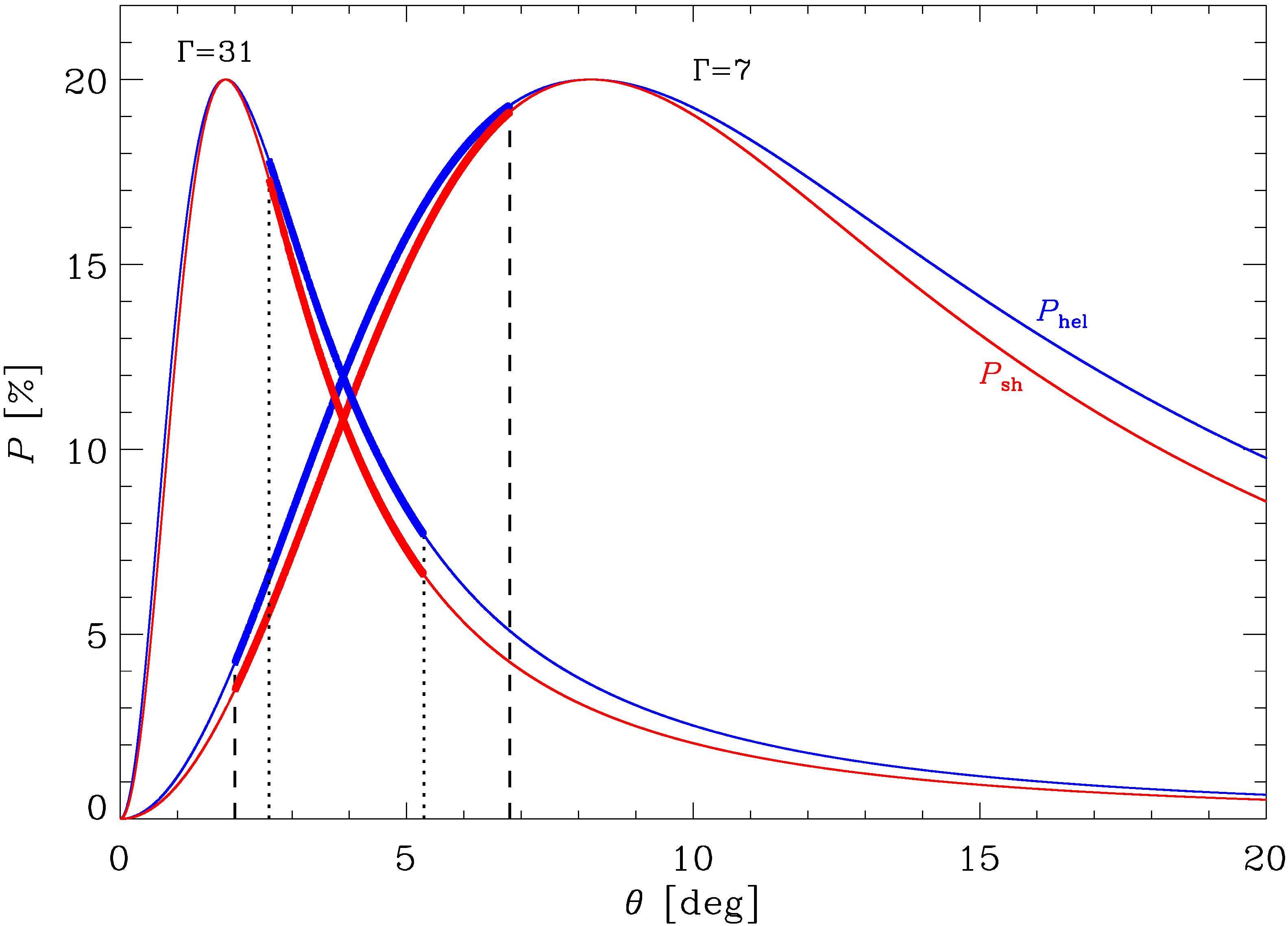}
\caption{Comparison of the behaviour of the polarisation degree versus the viewing angle  {of the optical emitting region} in 4C~38.41 and BL Lacertae. 4C~38.41 has a greater bulk Lorentz factor ($\Gamma=31$) than BL Lacertae ($\Gamma=7$), which makes the $P$ curve of 4C~38.41 reach a maximum for a viewing angle much smaller than in the BL Lacertae case. Therefore, for the range of $\theta$ (indicated by vertical lines) required to explain the long-term observed variability in the geometrical model, $P$ decreases for increasing $\theta$ (and thus for decreasing flux) in the 4C~38.41 case, while the opposite occurs for BL Lacertae. The blue and red lines refer to the helical magnetic field and shock-in-jet models, respectively. Figure adapted from \citep{raiteri2013}.\label{confronto}}

\end{figure}

In the same paper, the authors discussed the difference in polarisation behaviour of 4C~38.41 and BL Lacertae, i.e., an increase in $P_{\rm jet}$ with $F_{\rm jet}$ in the first case, and general anticorrelation in the second case. They explained this difference as the consequence of the value of the bulk Lorentz factor, which was assumed to be more than four times higher in 4C~38.41, implying a much stronger aberration of the viewing angle. Figure~\ref{confronto} shows the behaviour of $P$ as the observed viewing angle of the optical emitting region changes: in the case of 4C~38.41 the range of $\theta$ that explains the observed variability corresponds to a decreasing trend of $P$ when $\theta$ increases, i.e., when the flux decreases. In the case of BL Lacertae, the opposite occurs.

Seven years of multiwavelength data on the FSRQ OJ~248 were analysed by \cite{carnerero2015}, including polarisation and spectroscopy. As already found for 4C~38.41, $P$ was seen to increase with flux in this case as well. Moreover, the EVPA showed   both clockwise and anti-clockwise {large} rotations. The authors concluded that the emitting plasma is likely plunged in a turbulent magnetic field.

During the WEBT campaign on the BL Lac object S5~0716 + 714 in 2014, ref. \cite{bhatta2015} discovered a highly polarised optical microflare lasting about 5 h.
The authors separated the flaring component from the base-level one.
This allowed them to estimate the polarisation degree of the flaring component as $P$ $\sim$ $40$--60\%. The corresponding EVPA indicated a small misalignment with respect to the direction of the radio jet. A detailed analysis of the microflare properties suggested that it originated in a single jet region with a highly ordered magnetic field, a condition that can be  {provided by the compression in a shock or a sufficiently strong toroidal magnetic field component, for example.}

\begin{figure}[H]
\includegraphics[width=11.5 cm]{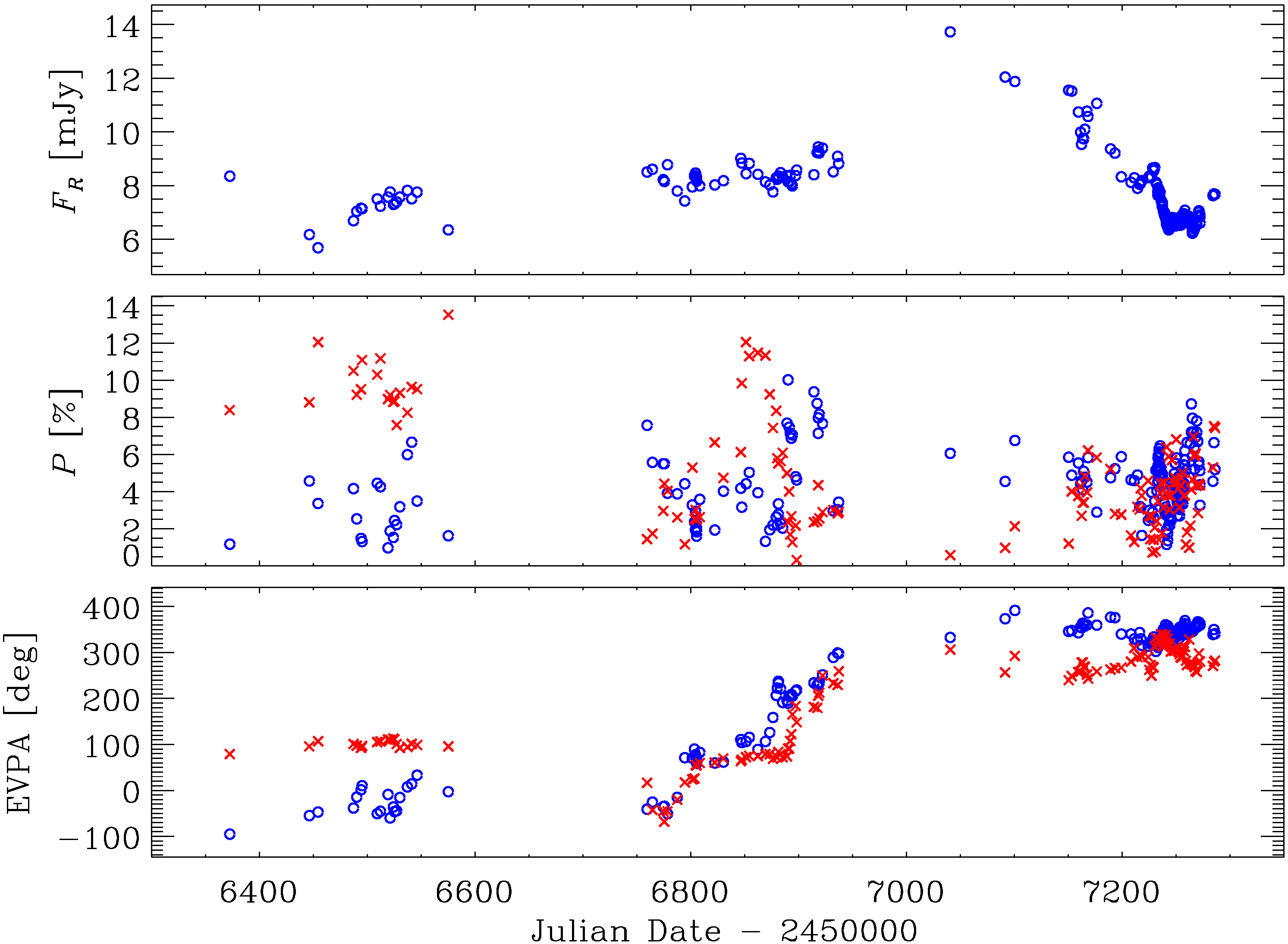}
\caption{(\textbf{Top}): The $R$-band flux densities of PG~1553 + 113 during the period 2013--2015. (\textbf{Middle}): The polarisation degree. (\textbf{Bottom}): The optical EVPA.  {The blue circles represent the observed quantities, while the red crosses refer to the result of the Monte Carlo simulations involving 220~emitting cells with a uniform and randomly oriented magnetic field.} Figure adapted from \citep{raiteri2017}.\label{stoka}}

\end{figure}

An analysis of the polarimetric behaviour of the BL Lac object PG~1553 + 113 was included in the paper by \cite{raiteri2017}, which reported the results of the WEBT monitoring of the source during the period 2013--2015, together with UV and X-ray observations by {\it XMM-Newton} and {\it Swift}. The value of $P$ went from about 1 to 10\% and did not correlate with flux.
The EVPA showed a  {large} rotation (more than $200^\circ$) in several months while the source was not particularly active. None of the deterministic models proposed for the interpretation of the blazar EVPA rotations was able to explain the observed behaviour; therefore, the authors investigated the stochastic hypothesis already explored by, e.g., \cite{marscher2014,blinov2015,kiehlmann2016}.
They run Monte Carlo simulations of the time evolution of the Stokes parameters in 220 emitting cells with randomly oriented uniform magnetic field. At each time step, the magnetic field in some of them was let to change its orientation, and the values of $P$ and EVPA were obtained by summing the Stokes parameters of all cells. The result was a set of $P$ and EVPA trends, one of which was particularly similar to the observed one, as shown in Figure~\ref{stoka}.

The bright and close BL Lac object Mkn~421 was studied by \cite{carnerero2017}. After correcting $P$ for the dilution effect by the host galaxy, it was found to vary between $\sim$$0.1\%$ and $\sim$$15\%$. The optical EVPA underwent  {large} rotations. The distribution of its values peaked at $\sim$$0^\circ$, similarly to that    {of the VLBA} 43 GHz  {radio core}. In contrast, the distribution of the radio EVPA at 15 GHz  {from the MOJAVE Program} indicated a prevalence of directions  {almost perpendicular to the previous ones}, suggesting that the low-frequency radio emission probably comes from a distinct jet region,  {possibly farther from the base of the jet}.
{This seems to be confirmed by the delay of about 45 days between the optical and 15 GHz flux density variations found by} \citep{acciari2021}.

The FSRQ CTA~102 remained in low activity from 2005 to 2012, when a noticeable outburst was observed. The multiwavelength source behaviour was analysed by \cite{larionov2016a}. The observed optical $P$ was seen to vary rapidly between 0\% and 12\% before the outburst, and to enlarge the range, up to $\sim$$20\%$, during and after the flaring state. In contrast,  {large} rotations of the optical EVPA were seen during the pre- and post-outburst phases, while more moderate EVPA swings occurred in the most dramatic outbursts phases.
All the strongest rotations were in the  {same (clockwise)} direction, which  {suggested that their origins were geometric, i.e., due to spiral motion. This} favoured a scenario involving a relativistic shock moving down a helical jet, or along helical magnetic field lines see, e.g.,~\citep[][]{marscher2010} and references therein.

After a few more years of modest activity, CTA~102 underwent an extraordinary outburst in 2016--2017.
In  \cite{raiteri2017_nature}, a detailed analysis of the optical light curves allowed to confirm that the long-term variability  {could be explained by variations} of the Doppler factor. Assuming that this {is most likely} produced by a change in the emitting jet orientation, the authors explained the optical--radio behaviour of the source in terms of a twisted inhomogeneous jet, where the emission at progressively increasing wavelength comes from zones progressively more distant from the jet apex. Modelling the big blue bump and summing its emission contribution to that of the jet, they were able to reproduce the SEDs of the source in various brightness states and to infer how the viewing angle changes with frequency, i.e.,\ along the jet. Moreover, they derived the trends in time of both the Doppler factor and the viewing angle for the optical and radio emissions.
When correcting for the variable Doppler effect, they found some flux flickering remains, which are likely the signature of intrinsic, energetic processes.
They also presented polarimetric data. After correcting for the dilution effect by the big blue bump, the jet optical polarisation degree showed strong variability, up to $\sim$$47\%$, but no correlation with the jet flux. The optical EVPA displayed  {large} rotations in both directions. The authors commented that this suggests the presence of turbulence. However, they also noticed that in correspondence with some flux peaks, the EVPA undergoes either a rapid change or an inversion of its rotation direction.
This was found in agreement with a rotating helical jet with a longitudinal magnetic field.

The analysis of the polarimetric behaviour of another FSRQ, 4C~71.07 \citep{raiteri2019}, revealed fast and ample changes of the jet polarisation degree, up to 47\%, with no correlation with flux. In this object too, the optical EVPA was seen to rotate both clockwise and counter-clockwise, favouring a stochastic process due to turbulence.

A new work on 3C~279 was  {recently} published by \cite{larionov2020}.
They interpreted the smoother and more limited variations of the radio $P$, as compared with the optical one, as the indication that the radio and optical emissions come from regions of different sizes, both pervaded by a turbulent magnetic field. The number of turbulent cells with random orientation in the radio zone must be larger than in the optical one, leading to a more destructive sum of components.
However, the optical and radio EVPA have roughly the same direction, mostly parallel to the jet axis. Therefore, in addition to turbulence, there must be a common mechanism that partially orders the magnetic field in both emitting regions, like shocks or a helical magnetic field component. Following \cite{larionov2016b}, the authors investigated the correlation between the normalised $u$ and $q$ Stokes parameters to verify the presence of monotonic EVPA rotations, which are revealed by a shift of the temporal evolution of one parameter with respect to that of the other. They found evidence of  {monotonic (anticlockwise)} rotation, which suggests a helical magnetic field component or a spiral structure of the jet,  {as in the above case of the clockwise rotations found in CTA~102 by} \citep{larionov2016a}.

\section{Conclusions}
Notwithstanding the strong observing effort spent to investigate the polarimetric properties of blazars and the huge theoretical work to interpret them, we still miss a definite picture. We see different behaviours in different objects, and also in the same object at different times. The strong scatter in the variations of the polarisation degree and EVPA, and the fact that EVPA rotations can occur also in non-flaring periods, and in both directions, likely imply that the magnetic field must have a turbulent component. On the other side, we sometimes see smooth EVPA rotations, or  rotations during flaring states, which suggest that some ordering mechanism must be present, as shock waves propagating in a helical jet, or along helical magnetic field lines.
The WEBT Collaboration will continue to investigate blazar variability in general, and polarimetric variability in particular, to provide further insight into this field.

\vspace{6pt}


\funding{This research received no external funding.}

\institutionalreview{Not applicable.}

\informedconsent{Not applicable.}

\dataavailability{The data collected by the WEBT collaboration are stored in the WEBT archive at the Osservatorio Astrofisico di Torino-INAF (\url{http://www.oato.inaf.it/blazars/webt/}); for questions regarding their availability, please contact the WEBT President Massimo Villata \mbox{(\url{massimo.villata@inaf.it})}.}


\conflictsofinterest{The authors declare no conflict of interest.}
\vspace{6pt}

\end{paracol}
\reftitle{References}

\end{document}